\documentclass[12pt,draftclsnofoot,onecolumn]{IEEEtran}
\usepackage{graphicx,cite,epsfig,amssymb,amsmath}

\begin{document}
\markboth{IEEE Transaction on Communications, Vol. XX,
No. Y, Month 2010} {}

\title{\mbox{}\vspace{1.00cm}\\
\textsc{Aggregate Interference Modeling in Cognitive Radio Networks with Power and Contention Control} \vspace{1.5cm}}

\author{\normalsize
Zengmao Chen$^{1}$, Cheng-Xiang Wang$^{1}$, Xuemin Hong$^{1}$, John Thompson$^{2}$\\
Sergiy A. Vorobyov$^{3}$, Xiaohu Ge$^{4}$, Hailin Xiao$^{5}$, and Feng Zhao$^{6}$\\
\vspace{0.7cm}
$^{1}$ Joint Research Institute for Signal and Image Processing\\
School of Engineering \& Physical Sciences\\
Heriot-Watt University, Edinburgh, EH14 4AS, UK.\\
Email:\{zc34, cheng-xiang.wang, x.hong\}@hw.ac.uk\\
\vspace{0.3cm}
$^{2}$Joint Research Institute for Signal and Image Processing\\
Institute for Digital Communications\\
University of Edinburgh,Edinburgh, EH9 3JL, UK.  \\
Email: john.thompson@ed.ac.uk\\
\vspace{0.3cm}
$^{3}$ Department of Electrical and Computer Engineering\\
University of Alberta, Edmonton, AB, T6G 2V4, Canada. \\
Email: vorobyov@ece.ualberta.ca\\
\vspace{0.3cm}
$^{4}$Department of Electronics and Information Engineering\\
Huazhong University of Science and Technology, Wuhan 430074, China.\\
Email: xhge@mail.hust.edu.cn\\
\vspace{0.3cm}
$^{5}$School of Information and Communication\\
Guilin University of Electronic Technology, Guilin 541004, China.\\
Email: xhl\_xiaohailin@yahoo.com.cn\\
\vspace{0.3cm}
$^{6}$Department of Science and Technology\\
Guilin University of Electronic Technology, Guilin 541004, China.\\
Email: zhaofeng@guet.edu.cn\\
}
\date{\today}
\renewcommand{\baselinestretch}{1.2}
\thispagestyle{empty} 
\maketitle
\thispagestyle{empty}
\newpage
\setcounter{page}{1}

\begin{abstract}

In this paper, we present an interference model for cognitive radio (CR) networks employing power control, contention control or hybrid power/contention control schemes. For the first case, a power control scheme is proposed to govern the transmission power of a CR node. 
For the second one, a contention control scheme at the media access control (MAC) layer, based on carrier sense multiple access with collision avoidance (CSMA/CA), is proposed to coordinate the operation of CR nodes with transmission requests.
The probability density functions of the interference received at a primary receiver from a CR network are first derived numerically for these two cases. For the hybrid case, where power and contention controls are jointly adopted by a CR node to govern its transmission, the interference is analyzed and compared with that of the first two schemes by simulations. Then, the interference distributions under the first two control schemes are fitted by log-normal distributions with greatly reduced complexity. Moreover, the effect of a hidden primary receiver on the interference experienced at the receiver is investigated. It is demonstrated that both power and contention controls are effective approaches to alleviate the interference caused by CR networks. Some in-depth analysis of the impact of key parameters on the interference of CR networks is given via numerical studies as well.\\


\end{abstract}

\newpage
\IEEEpeerreviewmaketitle


\section{Introduction}

With the requirement to improve spectrum utilization, the newly emerging cognitive radio (CR) technology \cite{cr1}--\!\cite{cr4} has attracted increasing attention. A CR network is envisioned to be capable of reusing the unused or underutilized spectra of incumbent systems (also known as primary networks) by sensing its surrounding environment and adapting its operational parameters autonomously. A CR system may coexist with a primary network on either an interference-free or interference-tolerant basis \cite{cr5}. For the former case, the CR system only exploits the unused spectra of the primary network, which consequently guarantees no interference to primary users. For the latter case, the CR system is allowed to share the spectra assigned to the primary network, under the condition that the CR network must not impose detrimental interference on the primary network. Therefore, modeling and analyzing the interference caused by CR networks is of great importance to reveal how the service of a primary network is deteriorated and how CR networks may be deployed.

In the literature, the existing research on interference modeling for CR networks mainly falls into three categories: spatial, frequency-domain and accumulated interference modeling. For spatial interference modeling, the fraction of white spaces available for CR networks was investigated in~\cite{white}~and~\cite{vt_magazine}. In~\cite{cellular}, the region of interference for CR receivers and region of communication for CR transmitters were studied for the case where a CR network coexists with a cellular network. The interference from CR devices to wireless microphones operating in TV bands was analyzed in~\cite{tv}, where the loss of reliable communication area of a wireless microphone due to the existence of CR devices was examined. CR interference in the frequency domain was also researched in the literature, e.g., the interference due to out-of-band emission of a wireless regional area network (WRAN) was analyzed in \cite{tv_gordon}.

As for accumulated interference modeling, in \cite{acc1}, the aggregate interference power from a sea of CR transmitters surrounding a primary receiver was derived. Also, the accumulated CR transmission power perceived at a primary receiver was given by integrating over the ``CR sea'' with a certain power density. The performance of a primary system was evaluated in \cite{acc2} in terms of outage probability caused by the interference from CR networks. The outage probability was derived for both underlay and overlay spectrum sharing cases. In \cite{acc3} the aggregate interference from multiple CR transmitters following a Poisson point process was approximated by a Gamma distribution and the probability of interference at a primary receiver was also given. It is worth noting that only pathloss was assumed for the interfering channel in \cite{acc1}--\!\cite{acc3}. Their work was extended by taking both shadowing and fading into account in \cite{xm} and \cite{menon}. Moreover, the probability density function (PDF) for accumulated interference and outage probability due to the aggregate interference from CR nodes were also derived in \cite{xm} and \cite{menon}, respectively.

However, in all the previous works \cite{white}--\!\cite{menon}, the CR transmitters were assumed to transmit at a fixed power level, i.e., no power control for CR transmitters was considered. Moreover, the CR nodes were all assumed to communicate with each other simultaneously. Thus, no contention control scheme was employed at the cognitive media access control (MAC) layer.
Some preliminary results on CR interference modeling were obtained in \cite{wcnc10} by incorporating either power or contention control scheme. In this paper, we extend the aggregate interference modeling in the following aspects. Firstly, a more realistic power control scheme than that in \cite{wcnc10} is proposed, and a hybrid power/contention control scheme is introduced. Secondly, the PDFs of interference perceived at a primary network from a CR network are derived numerically for the cases of power or contention control. The interference distribution of the hybrid control scheme is also analyzed and compared with that of the pure power control and pure contention control schemes by simulations. Furthermore, for the power and contention control schemes, their interference distributions are fitted by log-normal distributions, which greatly reduces computational complexity compared to a numerical approach to obtain PDFs. Finally, the impact of a hidden primary receiver on the aggregate interference is investigated for all the three schemes. The impact of several key parameters on the resulting interference is evaluated as well, which provides some insights for the deployment of CR networks.

The remainder of this paper is organized as follows. The system model is elaborated in Section~II. The detailed interference modeling is presented in Section~III. In Section~IV, the interference distributions are approximated by log-normal distributions. We incorporate the hidden primary receiver problem in Section~V.  The impact of several key parameters on the interference is analyzed via numerical studies in Section~VI. Finally, Section~VII concludes the paper.


\section{System Model}


The system model is illustrated in Fig. 1. It consists of a CR network coexisting with a primary transmitter-receiver pair. The active CR transmitters are distributed in a 2-dimensional plane outside the \textit{interference\ region} (IR) of the primary receiver as shown in Fig.~1. The IR is a disk centered at the primary receiver with a radius $R$. CR transmission is forbidden within this circular IR in order to protect the primary receiver against co-channel interference from the surrounding CR transmitters, since it is assumed that all the CR transmitters reside in the same frequency spectrum as the primary transmitter. We model the aggregate interference received at the primary receiver due to the existence of a CR network and investigate the impact of CR network deployment parameters on the resulting aggregate interference.

The underlying interference channels from CR transmitters to the primary receiver experience pathloss, shadowing and fading. The pathloss function $g(r_j)$ is
\begin{equation}\label{pathloss_eq}
g(r_j)=r_j^{-\beta}
\end{equation}
where $r_j$ is the distance between the $j$th $(j=1,2,\cdots)$ active CR transmitter and the primary receiver and $\beta$ is the pathloss exponent. The composite model for shadowing and fading can be expressed as the product of the long term shadowing and the short term multipath fading. In this paper, log-normal shadowing and Nakagami fading are considered. Let $h_j$ denote the channel gain for the composite shadowing and fading of the interference channel from the $j$th active CR transmitter to the primary receiver. The PDF of the composite channel gain $h_j$ can be approximated by the following log-normal distribution \cite{gordon}
\begin{equation}\label{eq2}
f_\mathrm{h}(x)\approx \frac{1}{\sqrt{2\pi}\sigma x} \mbox{exp}\left\{ -\frac{(\mbox{ln}(x) - \mu)^2}{2\sigma^2} \right\}
\end{equation}
where the mean $\mu$ and variance $\sigma^2$ can be expressed as
\begin{eqnarray}\label{eq3}
\mu&=&\left(\sum_{k=1}^{m-1}\frac{1}{k}-\mbox{ln}(m)-0.5772\right)+\mu_{\Omega}\ \\ \label{eq4}
\sigma^2&=&\sum_{k=0}^{\infty}\frac{1}{(m+k)^2} + \sigma_{\Omega}^2\
\end{eqnarray}
with $m$ standing for the Nakagami shape factor and $\mu_{\Omega}$ and $\sigma_{\Omega}^2$ denoting the standard mean and variance of the log-normal distribution, respectively.

Let $p_j$ denote the transmission power of the $j$th active CR transmitter. The accumulated power of the instantaneous interference received at the primary receiver can be expressed as
\begin{equation}\label{eq5}
Y=\sum_{j=1}^\infty p_jg(r_j)h_j.
\end{equation}
In this paper, we investigate the characteristics of the aggregate interference from all CR transmitters employing the following three different schemes: (i) power control, (ii) contention control, and (iii) hybrid power/contention control.

\subsection{Power Control}

In this scenario, the distribution of active CR transmitters follows a Poisson point process with a density parameter $\lambda$ for the density of CR transmitters on the plane.

The transmission power of a CR transmitter is governed by the following power control law
\begin{equation}\label{eq6}
p_\mathrm{pwc}(r_{\mathrm{cc}_j}) = \left\{
\begin{array}{ll}
\left(\frac{r_{\mathrm{cc}_j}}{r_\mathrm{pwc}}\right)^\alpha P_{\mathrm{max}},\ \ \ \ 0<r_{\mathrm{cc}_j}\leq r_\mathrm{pwc}\\
P_{\mathrm{max}},\ \ \ \ \ \ \ \ \ \ \ \ \ r_{\mathrm{cc}_j}>r_\mathrm{pwc}\\
\end{array} \right.
\end{equation}
where $r_{\mathrm{cc}_j}$ is the distance from the $j$th active CR transmitter to its nearest neighbouring active CR transmitter, $\alpha$ is the power control exponent, $P_{\mathrm{max}}$ is the maximum transmission power for CR transmitters, and $r_\mathrm{pwc}$ is the power control range, which determines the minimum $r_{\mathrm{cc}_j}$ leading to maximum CR transmission power $P_{\mathrm{max}}$. Compared to the power control law in \cite{wcnc10}, a new parameter $r_\mathrm{pwc}$ is introduced here to adjust the range of the power control. We assume that the power control exponent $\alpha$ is equal to the pathloss exponent $\beta$ in \eqref{pathloss_eq} throughout the paper. The above proposed power control scheme is designed in such a manner that the interference caused by the $j$th active CR transmitter to its nearest active CR transmitter due to pathloss is $p_\mathrm{pwc}(r_{\mathrm{cc}_j})g(r_{\mathrm{cc}_j})$. It is clear that within the power control range $r_\mathrm{pwc}$, this interference is equal to a constant $P_\mathrm{max}/r_\mathrm{pwc}^{\alpha}$. But beyond the power control range, the interference is less than that constant. In other words, at any CR transmitter the interference from the nearest neighbouring CR transmitter is capped and independent of the nearest neighbour distance within the power control range. It is worth noting that for each CR transmitter the information of its nearest neighbour distance is indispensable to determine its transmission power. Therefore, to facilitate the abovementioned power control scheme, either a central console having the global position information of all active CR transmitters or a distributed sensing scheme for CR transmitters like pilot sensing \cite{pilot_sensing} is required, the detail of which is, however, beyond the scope of this work. When CR transmitters follow a Poisson point distribution with a density $\lambda$, the PDF of $r_{\mathrm{cc}_j}$ can be given as \cite{stoyan}
\begin{equation}\label{eq7}
f_\mathrm{cc}(x)=2\pi \lambda x e^{-\lambda\pi x^2}.
\end{equation}

\subsection{Contention Control}

Unlike the previous power control scheme, for the case of contention control every active CR transmitter has fixed transmission power $p$, but their transmission is governed by contention control to determine which CR transmitters can transmit at a given time. We assume that the multiple access protocol carrier sense multiple access with collision avoidance (CSMA/CA) is employed, like in IEEE 802.11 networks. Every CR transmitter senses the medium before transmission. If the medium is busy, namely, the CR transmitter detects transmission from other CR transmitters within its contention region, it defers its transmission. Otherwise, the CR transmitter starts its transmission. As a result of the contention control shown, all the active CR transmitters are separated from each other by at least the contention distance, which is the minimum distance $d_{\mathrm{min}}$ between two concurrent CR transmitters.

The distribution of the active CR transmitters under the contention control can be modeled as a Matern-hardcore (MH) point process \cite{stochastic}, which can be considered as a thinned process from a Poisson point process \cite{stoyan}. The thinning operation deletes some points from the original Poisson process under certain criteria. The MH process $\Phi_\mathrm{mh}$ is the result of dependent thinning from a Poisson point process $\Phi$, i.e., deleting or retaining a point depends on previous deletion operations. The mathematical expression of the MH process is given by \cite{stoyan}
\begin{equation}\label{eq8}
\Phi_\mathrm{mh}\!\!=\!\!\{x\in\!\Phi\!:\!m(x)\!\!<\!m(y)\ \mbox{for\ all\ \em y}\mbox{\ in}\ \Phi\cap C(x,d_{\mathrm{min}})\}.
\end{equation}
Each point $x$ in the original Poisson point process $\Phi$ is marked with a random variable $m(x)$ uniformly distributed in (0,1), while $C(x,d_{\mathrm{min}})$ is a disk centered at point $x$ with the radius $d_{\mathrm{min}}$. The \textit{retaining\ probability} $q_{\mathrm{mh}}$ for the MH process, which is the probability of a point from a Poisson point process with a density $\lambda$ surviving the thinning process, is given by \cite{stoyan}
\begin{equation}\label{eq9}
q_{\mathrm{mh}}=\frac{1-e^{-\lambda \pi d_{\mathrm{min}}^2}}{\lambda \pi d_{\mathrm{min}}^2}.
\end{equation}

\subsection{Hybrid Power/Contention Control}

The aforementioned power control scheme regulates the transmission power of each CR transmitter according to its nearest neighbouring transmitter distance, while the contention control determines which CR transmitter can transmit at a time instant with fixed transmission power. A natural extension of the above two interference management schemes is to implement both schemes in the same system. This is termed hybrid power/contention control and it works in the following manner. The contention control scheme is first applied, resulting in a set of active CR transmitters following an MH point process. Then, a power control scheme similar to \eqref{eq6} is employed to adjust the transmission power of each active CR transmitter according to the distance to the nearest neighbouring active transmitter. The following power control law is adopted in the hybrid control scheme
\begin{equation}\label{new_power}
p_\mathrm{hyb}(r) = \left\{
\begin{array}{ll}
\left(\frac{r}{d_\mathrm{min}}\right)^\alpha p,\ \ \ \ d_\mathrm{min}\leq r\leq r_\mathrm{hyb}\\
\left(\frac{r_\mathrm{hyb}}{d_\mathrm{min}}\right)^\alpha p,\ \ \ \ \ \ \ \ \ \ \ \ r>r_\mathrm{hyb}
\end{array}\right.
\end{equation}
where $r$ is the distance from an active CR transmitter to its nearest neighbouring active CR transmitter, $\alpha$ is the power control exponent as in \eqref{eq6}, and $r_\mathrm{hyb}$ is the power control range similar to $r_\mathrm{pwc}$ in \eqref{eq6} except that it also determines the maximum transmission power, i.e., $\left(\frac{r_\mathrm{hyb}}{d_\mathrm{min}}\right)^\alpha p$. It is obvious that a larger $r_\mathrm{hyb}$ leads to a larger maximum CR transmission power and, consequently, longer communication range for CR transmitters. The above power control law \eqref{new_power} guarantees that when a pathloss channel is considered for each active CR transmitter, the perceived interference caused by its nearest neighbouring CR transmitter is $p_\mathrm{hyb}(r)g(r)$, which is (i) a constant $p/d^{\alpha}_\mathrm{min}$ within the power control range $r_\mathrm{hyb}$ and (ii) less than the constant $p/d^{\alpha}_\mathrm{min}$ when the distance $r$ is larger than the power control range.

\vspace{0.3cm}
\section{Interference Modeling}

We intend to model the aggregate interference from CR transmitters employing the three different interference management schemes introduced in Section II by finding their corresponding PDFs. We apply the methodology used, for example, in \cite{xm} and \cite{optimum90} to derive the PDFs.  First, the characteristic functions of the interference under different system models are derived. Then, the PDFs of the aggregate interference are obtained by performing an inverse Fourier transform on their characteristic functions.

\subsection{Power Control}
When all the CR transmitters follow a Poisson point process distribution and employ the power control scheme proposed in \eqref{eq6}, we can adopt the characteristic function-based method as in \cite{xm}, \cite{optimum90}-\cite{poisson06} and obtain the following characteristic function $\phi_\mathrm{Y}(\omega)$ of the aggregate interference $Y$ at a primary receiver from all CR transmitters
\begin{align}\label{eq10}
\phi_\mathrm{Y}(\omega)&=\mbox{exp}\left( \lambda \pi \int_H f_\mathrm{h}(h)\int_P f_\mathrm{p}(p) T(\omega p h) dp\ dh\right)
\end{align}
where $f_\mathrm{p}(\cdot)$ is the PDF of the transmission power $p_\mathrm{pwc}(r_{\mathrm{cc}_j})$ of a CR transmitter defined in \eqref{eq6} and
\begin{align}\label{eq11}
T(\omega p h)=R^2(1-e^{i\omega g(R)ph}) + i\omega ph\int^{g(R)}_0 \!\![g^{-1}(t)]^2e^{i\omega tph}dt.
\end{align}
In \eqref{eq11}, $g^{-1}(\cdot)$ denotes the inverse function of $g(\cdot)$ in \eqref{pathloss_eq}. For the derivation of \eqref{eq10}, the following fact is used: the distances from the $j$th CR transmitter to the primary receiver $r_j\ (j=1,2,\cdots)$ have independent and identical uniform distributions for a given number of CR transmitters~\cite{optimum90}. Their PDFs have the following form~\cite{optimum90}
\begin{equation}\label{eq12}
f_\mathrm{r}(x) = \left\{  \begin{array}{ll} 2x/(l^2-R^2), \ \ \ \ \ \ \ \ \ \ R \leq x \leq l \\
0,   \ \ \ \ \ \ \ \ \ \ \ \ \ \ \ \ \ \ \ \ \ \ \ \ {\mbox{otherwise}}
\end{array} \right.
\end{equation}
when CR transmitters are distributed within an annular ring with inner radius $R$ and outer radius $l$.
In \eqref{eq10}, $p$ is a function of $r_\mathrm{cc}$ as shown in \eqref{eq6}, so the expectation of $T(\omega p h)$ over $p$ equals that of $T(\omega p_\mathrm{pwc}(r_\mathrm{cc}) h)$ over $r_\mathrm{cc}$. Using the PDF of $r_\mathrm{cc}$ given in \eqref{eq7}, \eqref{eq10} can be rewritten as
\begin{align}\label{eq13}
\phi_\mathrm{Y}(\omega)=\mbox{exp}\left( \lambda \pi \int_H f_\mathrm{h}(h)\int_{r_\mathrm{cc}} f_\mathrm{cc}(r) T(\omega p_\mathrm{pwc}(r_\mathrm{cc}) h) drdh\right).
\end{align}
Moreover, \eqref{eq13} can be written as (see Appendix~A for the detailed derivation procedure)
\begin{align}\label{eq14}
&\phi_\mathrm{Y}\!(\omega)\!=
\!\mbox{exp}\!\left\{\!\lambda \pi \!\!\int_H\! f_\mathrm{h}\!(h)\!\int_0^{r_\mathrm{pwc}} \!\!f_\mathrm{cc}(r)\!\left[ R^2\!\!\left(1\!-\!e^{ \frac{i\omega r^\alpha P_\mathrm{max}g(R)h}{{r_\mathrm{pwc}}^{\alpha}}}\right) \right.\right. \nonumber\\
&\left.\hspace{5.8cm}\ \ +\  \frac{i\omega r^\alpha P_\mathrm{max}h}{{r_\mathrm{pwc}}^{\alpha}}\!\int^{g(R)}_0\!\!\!\!\! t^{-\frac{2}{\beta}}e^{\frac{i\omega t r^\alpha P_\mathrm{max}h}{{r_\mathrm{pwc}}^{\alpha}}}dt\right]\! dr\ \!\!dh \nonumber\\
&\hspace{1.5cm}\left.\ \ \ \ \ + \lambda \pi \!\!\!\int_H \!\!\!f_\mathrm{h}(h)\!\!\int_{r_\mathrm{pwc}}^\infty\!\!\!\! f_\mathrm{cc}(r)\!\!\left[ R^2\!\!\left(1-e^{i\omega g(R)P_\mathrm{max}h}\right) \!\!+\! i\omega P_\mathrm{max}h\!\!\int^{g(R)}_0\!\!\!\!\! t^{-\frac{2}{\beta}}e^{i\omega tP_\mathrm{max}h}dt\right] \!dr\ \!\!dh\!\right\}\!.
\end{align}

Finally, we obtain the PDF of the interference by performing the inverse Fourier transform on $\phi_\mathrm{Y}(\omega)$ as
\begin{align}\label{pdf}
f_\mathrm{Y}(y)=\frac{1}{2\pi}\int_{-\infty}^{+\infty}\phi_\mathrm{Y}(\omega)e^{-2\pi i \omega y}d\omega.
\end{align}

Equations \eqref{eq14} and \eqref{pdf} serve as general expressions for the characteristic function and PDF, respectively, of the interference under the power control scheme. As a special case, when the pathloss exponent $\beta=4$ and the radius of the interference region $R=0$, the PDF $f_Y(y)$ can be further simplified through similar steps to that used in \cite{optimum90} and obtained as
\begin{equation}\label{eq15}
f_\mathrm{Y}(y) = \frac{\pi}{2} K \lambda  y^{-3/2} \exp \left(- \frac{\pi^3 \lambda^2 K^2 }{4y} \right)
\end{equation}
where
\begin{equation}\label{eq16}
K=\sqrt{P_{\mathrm{max}}}\int_H f_\mathrm{h}(h)\sqrt{h}\ dh\ \left[ \int_0^{r_\mathrm{pwc}} 2\pi r \lambda e^{-\lambda\pi r^2} \left(\frac{r}{r_\mathrm{pwc}}\right)^{\frac{\alpha}{2}}dr +  e^{-\lambda\pi {r_\mathrm{pwc}}^2} \right].
\end{equation}
The detailed derivation procedure for $K$ can be found in Appendix~B.

\subsection{Contention Control}
As mentioned in Section~II.B, the distribution of CR transmitters can be modeled as an MH point process when the contention control is adopted. The MH process is a dependent thinning process from the original Poisson point process, which means that the positions of CR transmitters are correlated to each other. However, it is very difficult to obtain the distribution function like \eqref{eq12} for an MH point process in order to model the distance from an active CR transmitter to the primary receiver. Instead, we approximate the MH point process as an independent thinned Poisson point process with retaining probability $q_{\mathrm{mh}}$ given by \eqref{eq9}. Then, 
the transmission power for the $j$th CR transmitter is ${p}_j=\{0,p\}$, which is a random variable taking values $p$ or $0$ with probabilities $p_\mathrm{mh}$ and $1-p_\mathrm{mh}$, respectively. To this end, the contention control scheme can be interpreted as follows: all the CR transmitters still follow the original Poisson point process with intensity $\lambda$, but the $j$th CR transmitter has probability $q_{\mathrm{mh}}$ to transmit at power level $p$. The characteristic function of the accumulated interference can be found as
\begin{equation}\label{phi_y}
\phi_\mathrm{Y}(\omega)=\mathrm{exp}\left( \lambda \pi q_\mathrm{mh} \int_H f_\mathrm{h}(h) T(\omega p h)dh \right).
\end{equation}
The detailed derivation of \eqref{phi_y} is presented in Appendix~C.

Moreover, the PDF of the interference can be obtained from \eqref{phi_y} and \eqref{pdf}. As a special case, when no IR is implemented and the pathloss exponent $\beta=4$, this PDF can be simplified as \eqref{eq15}
with
\begin{equation}\label{eq22}
K = q_{\mathrm{mh}}\int_{H} f_{h}(h) \sqrt{ph}\ dh.
\end{equation}

It is worth noting that the approximation for the MH point process actually ignores the dependence among the CR transmitters and treats an MH point process as a result of independent thinning process from an original Poisson point process. The accuracy of this approximation is evaluated in Section~IV.

\subsection{Hybrid Power/Contention Control}

So far, the PDFs of the interferences received at a primary receiver from a CR network employing power control and contention control schemes have been derived. In order to model the aggregate interference under the hybrid control scheme, the nearest neighbouring distance distribution function analogous to \eqref{eq7} for active CR transmitters is indispensable to evaluate the transmission power designated in  \eqref{new_power}. Unfortunately, there is no closed-form expression for the nearest neighbour distance distribution function for an MH point process \cite{paloheimo}. Alternatively, several estimators have been used to statistically estimate the nearest neighbour distance distribution function in practice \cite{DStoyan}. However, statistical estimation is not practical for deriving the characteristic function in our case. Thus, we approach this problem numerically. 

The PDF for the aggregate interference under the hybrid control scheme is simulated in Fig.~2, where the interference PDFs for power and contention control are given as well for the purpose of comparison. It can be seen from this figure that both the mean and variance of the aggregate interference increase for the hybrid control scheme compared to either power or contention control schemes. However, the boosted interference is paid off by the increased CR communication area (coverage) for the hybrid control scheme. We define the coverage of each CR transmitter as a circular disk centered at a CR transmitter with radii being $\mathrm{min}({r}/{2},{r_\mathrm{pwc}}/{2})$, ${d_\mathrm{min}}/{2}$ and $\mathrm{min}({r}/{2},{r_\mathrm{hyb}}/{2})$ for power control, contention control and hybrid power/contention control schemes, respectively. Then, the received signal power at cell edge of a CR transmitter due to pathloss is ${2^\beta P_\mathrm{max}}/{{r_\mathrm{pwc}}^\beta}$, ${2^\beta p}/{d^{\beta}_\mathrm{min}}$ and ${2^\beta p}/{d^{\beta}_\mathrm{min}}$ for the above three aforementioned schemes. For the sake of comparison, let $r_\mathrm{pwc}=d_\mathrm{min}$ and $ P_\mathrm{max}=p$, which guarantees that the strength of the received signal power at cell edge of a CR transmitter is the same for all the three schemes. The overall coverage of the CR netwrok under different control schemes can be investigated numerically. With this setup, the overall coverage ratio for the power control, contention control and hybrid power/contention control is $1.0093,\ 1,\ \rm{and}\ 2.0229$, respectively. Two interesting facts are unveiled from this experiment. Firstly, the power control scheme leads to slightly smaller interference and slightly lager coverage compared to the contention control scheme, which suggests that power control is preferable to contention control in terms of lower resulting interference and larger coverage if the CR system can afford the complexity introduced by implementing the power control scheme. Secondly, the hybrid scheme tends to cause higher interference, but it greatly enlarges the coverage compared to power and contention control~schemes.

\section{Analytical Approximation}

In the previous section, to derive the PDFs for aggregate interference, the characteristic function-based method has been used which consists of two steps. Namely, characteristic function computation and Fourier transformation. This interference modeling approach is extremely computation-intensive, since generally closed-form expressions are not admitted for either step and the computations in both steps have to be performed numerically. It is desirable to model the aggregate interference with less complexity. An alternative approach to model the interference, which greatly reduces complexity, is to approximate interference  PDFs as certain known distributions. Observations from Fig.~2 suggest that the interference distribution for either power or contention control is positively skewed and heavy-tailed, which suggests a log-normal distribution. Thus, in this section, we fit the aggregate interference under power and contention control schemes to log-normal distributions. The theory behind the log-normal fitting is based on the following two facts. It has been shown that the sum of interference from uniformly distributed interferers in a circular area is asymptotically log-normal \cite{menon}, \cite{log-normal}. This ensures that the aggregate interference in these two schemes can be approximated as log-normal distributed. Meanwhile, the sum of randomly weighted log-normal variables can be modelled as a log-normal distribution as well~\cite{log-normal2}, which guarantees that the aggregate interference is still log-normal distributed even if the effect of shadow fading \eqref{eq2} is taken into account. In what follows, the log-normal fitting is performed using a cumulant-matching approach \cite{cumulant}, where the first two order cumulants of the aggregate interference $Y$ in \eqref{eq5} are used to estimate the mean and variance of the log-normal distribution function. Therefore, the exact PDFs of interference can be obtained. Fortunately, these cumulants have closed-form expressions for both control schemes. Consequently, it significantly reduces the complexity compared to the interference modeling carried out in Section~III. 

For the PDF of a log-normal variable $x$
\begin{equation}\label{logn_pdf}
p(x)=\frac{1}{\sqrt{2\pi}\sigma x} \mathrm{exp}\left( \frac{-(\mathrm{ln}(x)-\mu)^2}{2\sigma^2} \right)
\end{equation}
its mean $\mu$ and variance $\sigma^2$ can be estimated using its first two order cumulants $k_1$ and $k_2$ as follows \cite{logn_est}:
\begin{eqnarray}\label{mu_sigma}
\mu &=& \mathrm{ln} \frac{k_1}{\sqrt{\frac{k_2}{k^2_1}+1}}\\
\sigma^2 &=& \mathrm{ln}\left( \frac{k_2}{k^2_1}+1 \right).
\end{eqnarray}
In the context of interference distribution fitting, the $n$th cumulant $k_n$ of the aggregate interference $Y$ can be obtained from its characteristic function $\phi_\mathrm{Y}(\omega)$ via the following equation
\begin{equation}\label{cumulant_cal}
k_n=\frac{1}{i^n}\left[ \frac{\partial^n \mathrm{ln} \phi_\mathrm{Y}(\omega)}{\partial \omega^n} \right]_{\omega=0}.
\end{equation}

\subsection{Power Control}
From \eqref{eq14} and \eqref{cumulant_cal}, the cumulants for aggregate interference under the power control scheme can be derived as (see Appendix~D for detailed derivation)
\begin{align}\label{pwc_cum1}
k_n=&\frac{2\lambda\pi P_\mathrm{max}^n e^{n\mu + \frac{n^2\sigma^2}{2}}}{(n\beta -2)R^{n\beta-2}} \left[ \frac{n\alpha(n\alpha-2)\cdots2}{{r_\mathrm{pwc}}^{n\alpha}(2\pi\lambda)^\frac{n\alpha}{2}}\!\left(\! 1\!-\!e^{-\lambda\pi {r_\mathrm{pwc}}^2} \!\right) \right.\nonumber\\
&\quad \quad\quad\quad\quad\quad\quad\quad\left. - \sum_{i=1}^{\frac{n\alpha}{2}-1}\!\!\!\frac{n\alpha(n\alpha -2)\cdots (n\alpha -2i+2)}{(2\pi\lambda {r_\mathrm{pwc}}^2)^i}{r_\mathrm{pwc}}^{n\alpha -2i}e^{-\lambda \pi {r_\mathrm{pwc}}^2} \right].
\end{align}

To evaluate the accuracy of the log-normal approximation for the power control case, some comparisons are performed in Fig.~3(a). It can be seen from Fig.~3(a) that there is fairly good agreement between the interference PDFs derived in Section~III and the approximated counterparts. This approximation approach can be applied to both the pathloss-only and shadow fading channels.

\subsection{Contention Control}
Following the similar steps as in Appendix~D and given the characteristic function \eqref{phi_y} for the aggregate interference under contention control and also using \eqref{cumulant_cal}, we can find the $n$th cumulant $k_n$ of aggregate interference as
\begin{eqnarray}\label{ctc_cum}
k_n&=&\frac{\lambda \pi q_\mathrm{mh}}{i^n}  \int_H f_\mathrm{h}(h) \left[ -R^2\left(i p g(R)h \right)^n + n\left(i p h \right)^n \int^{g(R)}_0 t^{n-1-\frac{2}{\beta}}dt\right]dh\nonumber\\
&=&\lambda \pi q_\mathrm{mh} \left( \frac{n}{n-\frac{2}{\beta}}g^{n-\frac{2}{\beta}}(R) - R^2 g^n(R)  \right) p^n \int_H f_\mathrm{h}(h)h^n dh \nonumber\\
&=&\frac{2  p^n \left( 1-e^{-\lambda \pi d_\mathrm{min}^2} \right) e^{n\mu + \frac{n^2 \sigma^2}{2}}}{(n\beta -2) d_\mathrm{min}^2 R^{n\beta -2}}.
\end{eqnarray}

The accuracy evaluation of log-normal approximation under the contention control scheme is also performed and shown in Fig.~3(b). It can be seen from this figure that the log-normal approximation is fairly accurate compared to the simulated interference PDFs for either pathloss-only or shadow fading channels. Moreover, the derived interference PDF obtained from \eqref{pdf} and \eqref{phi_y} is validated against the simulated counterpart in Fig.~3(b) as well, which suggests that the approximation for the MH point process in the derivation is reasonable.

\section{Imperfect Primary System Knowledge}
In practice, some information about the primary system may not be perfectly known. One prominent example is the location of the primary receivers, which is usually required by CR networks in order to protect primary receivers from interfering CR transmitters. However, this information is not always available, especially in the case of passive primary receivers, i.e., when the primary receivers are hidden from CR networks. It is widely accepted that passive receiver detection techniques can be used or developed in the context of CR networks. For example, one of such primary receiver detection techniques is reported in \cite{primary_detection}. Nevertheless, its applicability is still not convincingly viable since it requires deploying sensor nodes close to primary receivers and much coordination is involved between these sensors and CR networks as well. The most commonly used and also the simplest approach to protect the primary receiver is to regulate the transmission of the CR network based on primary transmitter sensing, assuming that primary receivers are in close proximity to the primary transmitter. In this section, we evaluate the effect of a hidden primary receiver on the resulting interference to primary receivers.

Consider a primary and CR coexisting systems depicted in Fig.~4, where an IR with radius $R$ centered at the primary transmitter is introduced. All CR transmitters are distributed in the shaded concentric ring with inner radius $R$ and outer radius $l$. Let $\theta$ be the angle between the line joining the primary receiver and a CR transmitter and the line joining the primary transmitter-receiver pair. The distance from the CR transmitter to the primary transmitter is $r$ and the distance between the primary transmitter-receiver pair is $r_\mathrm{p}$. Then, the distance between the CR transmitter and the primary receiver $r_\mathrm{cp}$ can be expressed as
\begin{equation}\label{cp_distance}
r_\mathrm{cp}(r,\theta)=r_\mathrm{p}\mathrm{cos}(\theta) + r \mathrm{sin} \left( \mathrm{cos}^{-1} \frac{r_\mathrm{p}\mathrm{sin}(\theta)}{r} \right),\ \ \  r\in [R, l];\  \theta \in [0, 2\pi]
\end{equation}
where $r$ is distributed as in \eqref{eq12} and $\theta$ is uniformly distributed in $[0, 2\pi]$ if a Poisson point process is assumed for the CR transmitter distribution.

\subsection{Power Control}
Under the power control scheme proposed in Section~II.A and the system model given in Fig.~4, the characteristic function of aggregate interference $\phi_\mathrm{Y}(\omega)$ can be written as follows (see Appendix~E for the detailed derivation):
\begin{align}\label{pwc_imp}
\phi_\mathrm{Y}(\omega)\!
&=\!\lim_{l\to \infty} \mathrm{exp}\! \left\{\lambda  \int_H \! f_\mathrm{h}(h)\!\int_0^{r_\mathrm{pwc}}\!\! f_\mathrm{cc}(x)\! \int_0^{2\pi}\!\!  \int_R^{l} 
e^{i\omega \left( \frac{r}{r_\mathrm{pwc}} \right)^\alpha P_\mathrm{max}(x) g(r_\mathrm{cp}(r,\theta)) h }
 r \!- r\ dr \ \!d\theta \ \!dx\ \!dh \right. \nonumber\\
&\ \ \ \ \ \ \ \ \ \ \ \  + \left. \lambda  \!\int_H \! f_\mathrm{h}(h)\!\int_{r_\mathrm{pwc}}^{\infty} f_\mathrm{cc}(x) \int_0^{2\pi} \!\! \int_R^{l} e^{i\omega  P_\mathrm{max}(x) g(r_\mathrm{cp}(r,\theta)) h} r - r\  dr \ \!d\theta \ \!dx\ \!dh \right\}.
\end{align}

Applying the log-normal approximation method used in Section~IV, we obtain the $k$th cumulant of interference as
\begin{align}\label{cum_pwc_imp}
k_n&=\lim_{l\to \infty}\lambda \left\{\!\int_H \! f_\mathrm{h}(h)\!\int_0^{r_\mathrm{pwc}} f_\mathrm{cc}(x) \int_0^{2\pi}  \int_R^{l} \frac{\left(r^\alpha P_\mathrm{max}(x) g(r_\mathrm{cp}(r,\theta)) h\right)^n}{{r_\mathrm{pwc}}^{n\alpha}}  r dr\ \! d\theta \ \! dx\ \! dh \right.\nonumber\\
&\ \ \ \ \ \ \ \ \ \  \left. + \!\int_H \! f_\mathrm{h}(h)\!\int_{r_\mathrm{pwc}}^\infty f_\mathrm{cc}(x) \int_0^{2\pi}  \int_R^{l} \left[ P_\mathrm{max}(x) g(r_\mathrm{cp}(r,\theta)) h\right]^n r dr\ \! d\theta \ \! dx\ \! dh \right\}.
\end{align}
As can be seen from \eqref{cum_pwc_imp}, unlike \eqref{pwc_cum1}, the $k$th cumulant does not have a closed-form expression. However, the complexity of obtaining the exact interference PDF from \eqref{cum_pwc_imp} is still smaller than that of the numerical method in Section III.

An experiment is performed in Fig.~5(a) to examine the effect of hidden primary receiver on the resulting interference compared to the interference for the case of perfect knowledge of primary receiver location. We consider a pathloss-only channel in this figure. It can be seen from the figure that the hidden primary receiver problem boosts the interference in terms of increased interference mean and variance. This figure also shows that the log-normal approximation still fits well the interference distribution in this scenario.

\subsection{Contention Control}
Under the contention control scheme proposed in Section~II.B and the system model given in Fig.~4, the characteristic function of aggregate interference $\phi_Y(\omega)$ can be expressed as
\begin{align}\label{ctc_imp}
\phi_Y(\omega)&= \lim_{l\to \infty} \mathrm{exp}\left\{q_\mathrm{mh}\lambda \pi D_l \left( E\left( e^{i\omega p g(V)h} \right) -1 \right)\right\}\nonumber\\
&=\lim_{l\to \infty} \mathrm{exp} \left\{q_\mathrm{mh} \lambda \pi D_l \left( \!\int_H \! f_\mathrm{h}(h)\!\int_0^{2\pi} \frac{1}{2\pi} \int_R^{l} \mathrm{exp}\left[i\omega p g(r_\mathrm{cp}(r,\theta)) h\right] \frac{2r}{D_l} dr \ \!d\theta\ \!dh  -1 \right) \right\}\nonumber\\
&=\lim_{l\to \infty} \mathrm{exp} \left\{q_\mathrm{mh} \lambda  \!\int_H \! f_\mathrm{h}(h)\!\int_0^{2\pi} \int_R^{l} \mathrm{exp}\left[i\omega p g(r_\mathrm{cp}(r,\theta)) h\right] r -r dr \ \!d\theta\ \!dh  \right\},
\end{align}
with $D_l=l^2-R^2$.

Using the same log-normal approximation method as in Section~IV, the $k$th cumulant of interference can be written as
\begin{align}\label{cum_ctc_imp}
k_n&=\lim_{l\to \infty} q_\mathrm{mh} \lambda \!\int_H \! f_\mathrm{h}(h)\!\int_0^{2\pi} \int_R^{l} \left[p g(r_\mathrm{cp}(r,\theta)) h\right]^n r -r dr \ \!d\theta \ \!dh.
\end{align}

The effect of hidden primary receiver under contention control is evaluated in Fig.~5(b), where a pathloss-only channel is assumed. As we can see from this figure, the uncertainty about the primary receiver location leads to interference with larger mean and variance as compared to that in the case with perfect knowledge of primary receiver location. Moreover, it can be seen from this figure that the log-normal fitting for the interference is fairly accurate and the approximation approach is still applicable in this scenario.

For the case of hybrid power/contention control, the effect of hidden priamry receiver cannot be examined analytically because the closed-form interference PDF is not available. Therefore, it is analyzed numerically in Fig.~6, whose initial setup is the same as the one used in Fig.~5(b) except that the power control range is $r_\mathrm{hyb}=30$~m. It can bee seen from Fig.~6 that the uncertainty about the primary receiver location boosts the interference in terms of increased mean, variance, and heavier tails for the hybrid control scheme as well. More interestingly, another two facts can be found by comparing Figs.~5 and 6: (i) the hidden primary receiver phenomonon has similar impact on the pure power and pure contention control schemes; (ii) the hybrid power/contention control scheme is less sensible to the phenomenon of hidden primary receiver than any of the other two schemes.
\section{Numerical Studies}

The aggregate interference power from CR transmitters employing power control or contention control is investigated numerically in this section. For the power control scheme, Fig.~7(a) shows the effect of different power control parameters on their resulting aggregate interference. The detailed setup for the initial power control scheme is as follows: the maximum transmission power for each CR transmitter $P_{\mathrm{max}}=1$ W, the density of CR transmitter $\lambda=3$~user/10$^4$m$^2$, the IR radius $R=100$ m, the power control range $r_\mathrm{pwc}=20$ m, the pathloss exponent $\beta=4$ and the power control exponent $\alpha=4$. From the two rightmost PDFs in this figure, it can seen that introducing power control scheme actually shifts the interference distribution leftwards compared to the distribution without power control. It means that the power control scheme can reduce the interference experienced at the primary receiver in terms of reducing its mean and slightly decreasing its variance. When deploying a CR network under the power control scheme, its resulting interference can be controlled by manipulating several parameters including $P_{\mathrm{max}}$, $r_\mathrm{pwc}$, $\lambda$, and $R$. It can be seen in Fig.~7(a) that the interference can be reduced by either decreasing the maximum transmission power and/or CR density, or increasing the power control range and/or IR radius. Interestingly, it also suggests that adjusting the IR radius is an effective way to control the interference, since the interference is more sensitive to the IR radius than to any other parameter as demonstrated in Fig.~7(a). Meanwhile, the interference is least sensitive to the CR user density in the sense that halving $\lambda$ leads to higher interference compared to doubling $r_\mathrm{pwc}$, halving $P_{\mathrm{max}}$ or doubling $R$.

For the contention control scheme, the impact of contention control parameters on the resulting interference is depicted in Fig.~7(b), whose initial setup is the same as that of Fig.~7(a) except that the transmission power for each CR transmitter is $p=1$~W and the contention control range is $d_\mathrm{min}=20$~m. It can be seen from the two rightmost PDFs in Fig.~7(b) that the contention control scheme results in an interference distribution with reduced mean like the power control scheme in Fig.~7(a). Meanwhile, the interference can be reduced by decreasing $p$, $\lambda$, and/or increasing $R$ or $d_\mathrm{min}$. It can be observed by comparing Fig.~7(b) with Fig.~7(a) that (i) increasing the IR radius is an effective approach to reduce the interference for both the power and contention control schemes. However, the power control scheme is more sensitive to the IR radius than the contention control one; (ii) reducing the transmission power and/or CR transmitter density affects the interference in the very similar manner for these two control schemes.

Finally, the impact of shadow fading on the aggregate interference is investigated for different values of the Nakagami shaping factor $m$ under power and contention control schemes, respectively, in Figs.~8(a) and 8(b).  The initial setup in this example is the same as the one used for Figs.~7(a) and 7(b), except that the standard variance is $\sigma_\Omega=4$ dB. When $m=1$ the interfering channel becomes a Rayleigh channel, which is dominated by the log-normal shadowing. Whereas, when $m=100$ the fluctuations of the channel are reduced significantly compared to the Rayleigh fading channel. One fact observed in Fig.~8 is that the interference distributions have larger variance and heavier tails when shadow fading is incorporated for both control schemes. Interestingly, fading tends to make the interference distribution more heavy-tailed than shadowing, i.e., the interference under shadowing has better outage property than that under fading. Moreover, the shadow fading has the similar effect for both control schemes. 

\section{Conclusions}

Interference at a primary receiver caused by CR transmitters with power control, contention control, and hybrid power/contention control schemes has been characterized. The PDFs of interference in the first two cases have been evaluated analytically while, the interference distribution under the hybrid power/contention control has been studied numerically. It has been found that the proposed power control and contention control schemes are two effective approaches to alleviate interference caused by CR transmitters. The hybrid control scheme causes higher interference to a primary receiver, but leads to larger CR coverage as compared to either power or contention control schemes. Then, the interference distributions for power and contention control schemes have been approximated by log-normal distributions using the cumulant-matching approach where the interference PDFs have been obtained with reduced complexity. Furthermore, the effect of a hidden primary receiver on the perceived interference has also been investigated for the primary receiver. Numerical studies have demonstrated the impact of some CR deployment parameters on the resulting aggregate interference under power and contention control schemes. It has been shown that increasing the IR radius is an effective way to reduce the interference. Moreover, the power control scheme is more sensitive to the IR radius than the contention control counterpart. Finally, the impact of shadow fading on the aggregate interference has been analyzed as well. 

\begin{center}
\textsc{{Acknowledgments}}\\
\end{center}

{\small Z. Chen, C.-X. Wang, X. Hong, and J. Thompson acknowledge the support from the Scottish Funding Council for the Joint Research Institute in Signal and Image Processing between the University of Edinburgh and Heriot-Watt University, as a part of the Edinburgh Research Partnership in Engineering and Mathematics (ERPem). S. A. Vorobyov acknowledges the support in part from the Natural Sciences and Engineering Research Council (NSERC) of Canada and in part from the Alberta Ingenuity Foundation, Alberta, Canada. X. Ge acknowledges the support from National Natural Science Foundation of China (NSFC) (Grant No.: 60872007), National 863 High Technology Program of China (Grant No.: 2009AA01Z239), and the Ministry of Science and Technology (MOST) of China, International Science and Technology Collaboration Program (Grant No.: 0903). F. Zhao acknowledges the support from the NSFC (Grant No.: 60872022). C.-X. Wang and F. Zhao acknowledge the support of the Key Laboratory of Cognitive Radio and Information Processing (Guilin University of Electronic Technology), Ministry of Education, China. The authors acknowledge the support from the RCUK for the UK-China Science Bridges Project: R\&D on (B)4G Wireless Mobile Communications.}

\begin{center}
\textsc{{Appendix}}\\
\end{center}
\emph{A. Derivation of \eqref{eq14}}

Substituting \eqref{eq6} and \eqref{eq7} into \eqref{eq13}, we have
\vspace{-0.7cm}
{\small\begin{align}\label{der_eq14}
&\phi_\mathrm{Y}\!(\omega)\!=\!\mbox{exp}\!\left\{\!\lambda \pi \!\int_H \!f_\mathrm{h}(h)\!\int_{r_\mathrm{cc}} \!f_\mathrm{cc}(r)\!\left[R^2\!\left(1\!-\!e^{i\omega g(R)p(r)h}\right) \!+\! i\omega p_\mathrm{pwc}(r_\mathrm{cc})h\!\int^{g(R)}_0\!\!\!\!\!{(g^{-1}(t))}^2e^{i\omega tp(r)h}dt\right]\! dr\ \! dh\right\} \nonumber\\
&=\mbox{exp}\!\left\{\!\!\lambda \pi \!\!\int_H\!\! f_\mathrm{h}(h)\!\int_0^{r_\mathrm{pwc}}\!\!\!\!\!\!\!\! \!f_\mathrm{cc}(r)\!\left[\! R^2\!\left(1\!-\!e^{i\omega (\frac{r}{r_\mathrm{pwc}})^{\alpha}P_\mathrm{max}g(R)h}\right)\! + \! \frac{i\omega r^\alpha P_\mathrm{max}h}{r_\mathrm{pwc}^\alpha}\!\int^{g(R)}_0\!\!\!\!\!{(g^{-1}(t))}^2e^{i\omega t(\frac{r}{r_\mathrm{pwc}})^{\alpha}P_\mathrm{max}h}dt\right] \!dr\ \!\!dh \right.\nonumber\\
&\hspace*{1cm}\left.+\! \lambda \pi \!\int_H \!f_\mathrm{h}(h)\int_{r_\mathrm{pwc}}^\infty f_\mathrm{cc}(r)\left[ R^2\left(1-e^{i\omega g(R)P_\mathrm{max}h}\right) + i\omega P_\mathrm{max}h\int^{g(R)}_0\!\!\!{(g^{-1}(t))}^2e^{i\omega tP_\mathrm{max}h}dt\right]\! dr\ \!dh\right\}.
\end{align}}
Using \eqref{pathloss_eq} and \eqref{der_eq14}, the characteristic function \eqref{eq14} is obtained.

\vspace{0.5cm}
\emph{B. Derivation of \eqref{eq16}}
{\small \begin{align}\label{der_eq16}
K&=\int_H f_\mathrm{h}(h)\int_P f_\mathrm{p}(p)\sqrt{hp}\ dp\ \! dh\nonumber\\
&=\sqrt{P_{\mathrm{max}}}\int_H f_\mathrm{h}(h)\sqrt{h}\ dh\ \left( \int_0^{r_\mathrm{pwc}} 2\pi r \lambda e^{-\lambda\pi r^2} \left(\frac{r}{r_\mathrm{pwc}}\right)^{\frac{\alpha}{2}}dr +  \int_{r_\mathrm{pwc}}^\infty 2\pi\lambda r e^{-\lambda\pi r^2} \ dr \right),
\end{align}}
where the first equality of \eqref{der_eq16} holds according to \cite{optimum90}. \eqref{eq16} is obatined immediately from \eqref{der_eq16}.

\vspace{0.5cm}
\emph{C. Derivation of \eqref{phi_y}}

Following similar steps as in \cite{xm}, the characteristic function of the aggregate interference can be expressed as
\begin{equation}\label{eq18}
\phi_\mathrm{Y}(\omega)=\lim_{l\to \infty}e^{\lambda \pi (l^2-R^2)(Q-1)}
\end{equation}
where
{\small\begin{align}\label{q_eq}
Q&=E\left( e^{i\omega P g(V) H}\right)\nonumber\\
&=\int_H f_\mathrm{h}(h) \int^{l}_{R} E\left[ e^{i\omega P g(r) h} \right] \frac{2r}{l^2-R^2} \ dr \ dh\nonumber\\
&=\int_H f_\mathrm{h}(h) \int^{l}_{R} \left[(1-q_\mathrm{mh}) + q_\mathrm{mh}e^{i\omega p g(r) h}\right] \frac{2r}{l^2-R^2} \ dr \ dh \nonumber\\
&=1-q_\mathrm{mh} + q_\mathrm{mh} \int_H f_\mathrm{h}(h)\int^{l}_{R} e^{i\omega p g(r) h} \frac{2r}{l^2-R^2} \ dr \ dh.
\end{align}}
The integral in the last equality of \eqref{q_eq} can be written as
\begin{equation}\label{integral}
\lim_{l\to \infty} \int_H f_\mathrm{h}(h)\int^{l}_{R} e^{i\omega p g(r) h} \frac{2r}{l^2-R^2} \ dr \ dh 
=1+\frac{1}{l^2-R^2}\int_H f_\mathrm{h}(h) T(\omega p h)dh
\end{equation}
where $T(\omega p h)$ is given in \eqref{eq11}. Substituting \eqref{q_eq} and \eqref{integral} into \eqref{eq18}, we obtain \eqref{phi_y}.

\vspace{0.5cm}
\emph{D. Derivation of \eqref{pwc_cum1}}

From \eqref{eq14} and \eqref{cumulant_cal}, we have
\vspace{0.5cm}
{\small\begin{align}\label{pwc_cum}
\!\!k_n\!\!&=\!\!\frac{\lambda \pi}{i^n}\!\int_H \!f_\mathrm{h}(h) \!\int^{r_\mathrm{pwc}}_0 \!f_\mathrm{cc}(r)\!\left[ -R^2\left(\frac{i r^\alpha \!P_\mathrm{max} g(R)h}{{r_\mathrm{pwc}}^\alpha}  \right)^n \!+\! \frac{n\left(i r^\alpha P_\mathrm{max} h \right)^n}{{r_\mathrm{pwc}}^{n\alpha}}  \!\int^{g(R)}_0 \!t^{n-1-\frac{2}{\beta}}dt\right]dr\ \!dh\nonumber\\
\!\!&\ +\frac{\lambda \pi}{i^n}\int_H f_\mathrm{h}(h) \int^\infty_{r_\mathrm{pwc}} f_\mathrm{cc}(r)\left[ -R^2\left(i P_\mathrm{max} g(R)h \right)^n + n\left(i P_\mathrm{max} h \right)^n \int^{g(R)}_0 t^{n-1-\frac{2}{\beta}}dt\right]dr\ \!dh\nonumber\\
\!\!&=\!\!\lambda \pi \!\int_H\! \!f_\mathrm{h}(h)h^n dh \left( \frac{n}{n-\frac{2}{\beta}}g^{n-\frac{2}{\beta}}(R) \!- \!R^2 g^n(R) \right)\! \left[ \int^{r_\mathrm{pwc}}_0 \!f_\mathrm{cc}(r)  \frac{(r^\alpha P_\mathrm{max})^n}{{r_\mathrm{pwc}}^{n\alpha}}   \!dr \!+\! \int^\infty_{r_\mathrm{pwc}} \!\!\!f_\mathrm{cc}(r) P_\mathrm{max}^n dr \!\right]\nonumber\\
\!\!&=\!\frac{2\lambda\pi P_\mathrm{max}^n}{(n\beta -2)R^{n\beta-2}} \int_H f_\mathrm{h}(h)h^n dh \left( \int^{r_\mathrm{pwc}}_0 \frac{f_\mathrm{cc}(r) r^{n\alpha}}{{r_\mathrm{pwc}}^{n\alpha}}  dr + \int^\infty_{r_\mathrm{pwc}} f_\mathrm{cc}(r) dr \right).
\end{align}}
The first equality of \eqref{pwc_cum} is obtained based on the following fact
\begin{equation}\label{partial}
\left[\frac{\partial^n}{\partial \omega^n}\right]_{\omega=0} e^{a \omega} 
= \left[\frac{\partial^n}{\partial \omega^n}\right]_{\omega=0} \sum_{i=0}^{\infty} \frac{(a \omega)^n}{n!}
=a^n.
\end{equation}
In the last equality of \eqref{pwc_cum}, the first integral can be expressed as \cite{logn_dis}
\begin{equation}\label{logn_int}
\int_H f_\mathrm{h}(h)h^n dh=e^{n\mu + \frac{n^2\sigma^2}{2}}
\end{equation}
with $\mu$ and $\sigma^2$ given in \eqref{eq3} and \eqref{eq4}, respectively. Also, the sum of the last two integrals in \eqref{pwc_cum} can be simplified as
{\small\begin{align}\label{int}
&\int^{r_\mathrm{pwc}}_0 \frac{f_\mathrm{cc}(r) r^{n\alpha}}{{r_\mathrm{pwc}}^{n\alpha}}  dr + \int^\infty_{r_\mathrm{pwc}} f_\mathrm{cc}(r) dr\nonumber\\
&=\frac{n\alpha(n\alpha-2)\cdots2}{{r_\mathrm{pwc}}^{n\alpha}(2\pi\lambda)^\frac{n\alpha}{2}}\left( 1-e^{-\lambda\pi {r_\mathrm{pwc}}^2} \right) - \sum_{i=1}^{\frac{n\alpha}{2}-1}\frac{n\alpha(n\alpha -2)\cdots (n\alpha -2i+2)}{(2\pi\lambda {r_\mathrm{pwc}}^2)^i}{r_\mathrm{pwc}}^{n\alpha -2i}e^{-\lambda \pi {r_\mathrm{pwc}}^2}.
\end{align}}
Substituting \eqref{logn_int} and \eqref{int} into \eqref{pwc_cum} yields \eqref{pwc_cum1}.

\vspace{0.5cm}
\emph{E. Derivation of \eqref{pwc_imp}}
{\small\begin{align}\label{der_pwc_imp}
\!\!\!\phi_\mathrm{Y}(\omega)&= \!\lim_{l\to \infty} \mathrm{exp}\left\{\lambda \pi D_l \left( E\left( e^{i\omega p_\mathrm{pwc} g(V)h} \right)\! -\!1 \!\right)\!\right\}\nonumber\\
&=\!\lim_{l\to \infty}\! \mathrm{exp}\! \left\{\!\lambda \pi D_l \!\left[ \!\int_H \! f_\mathrm{h}(h)\!\int_0^{\infty}\!\!\! f_\mathrm{cc}(x)\!\! \int_0^{2\pi}\!\!\! \frac{1}{2\pi} \!\int_R^{l} \mathrm{exp}\left[i\omega p_\mathrm{pwc}(x) g(r_\mathrm{cp}(r,\theta)) h\right] \frac{2r}{D_l} dr \ \!d\theta \ \!dx\ \!dh \!\! -\!1 \right] \right\}\nonumber\\
&=\lim_{l\to \infty} \mathrm{exp} \left\{\lambda  \!\int_H \! f_\mathrm{h}(h)\!\int_0^{\infty} f_\mathrm{cc}(x) \int_0^{2\pi}  \int_R^{l} \mathrm{exp}\left[i\omega p_\mathrm{pwc}(x) g(r_\mathrm{cp}(r,\theta)) h\right] r - r dr \ \!d\theta \ \!dx\ \!dh \right\}
\end{align}}
with $D_l=l^2-R^2$. The first equality in \eqref{der_pwc_imp} is obtained in the same way as \eqref{eq18} and \eqref{q_eq}. Equation \eqref{pwc_imp} can be obtained immediately from \eqref{der_pwc_imp}.


\newpage
\begin{center}
\begin{tabular}{c}
\hskip -0.6cm\epsfxsize=13cm\epsffile{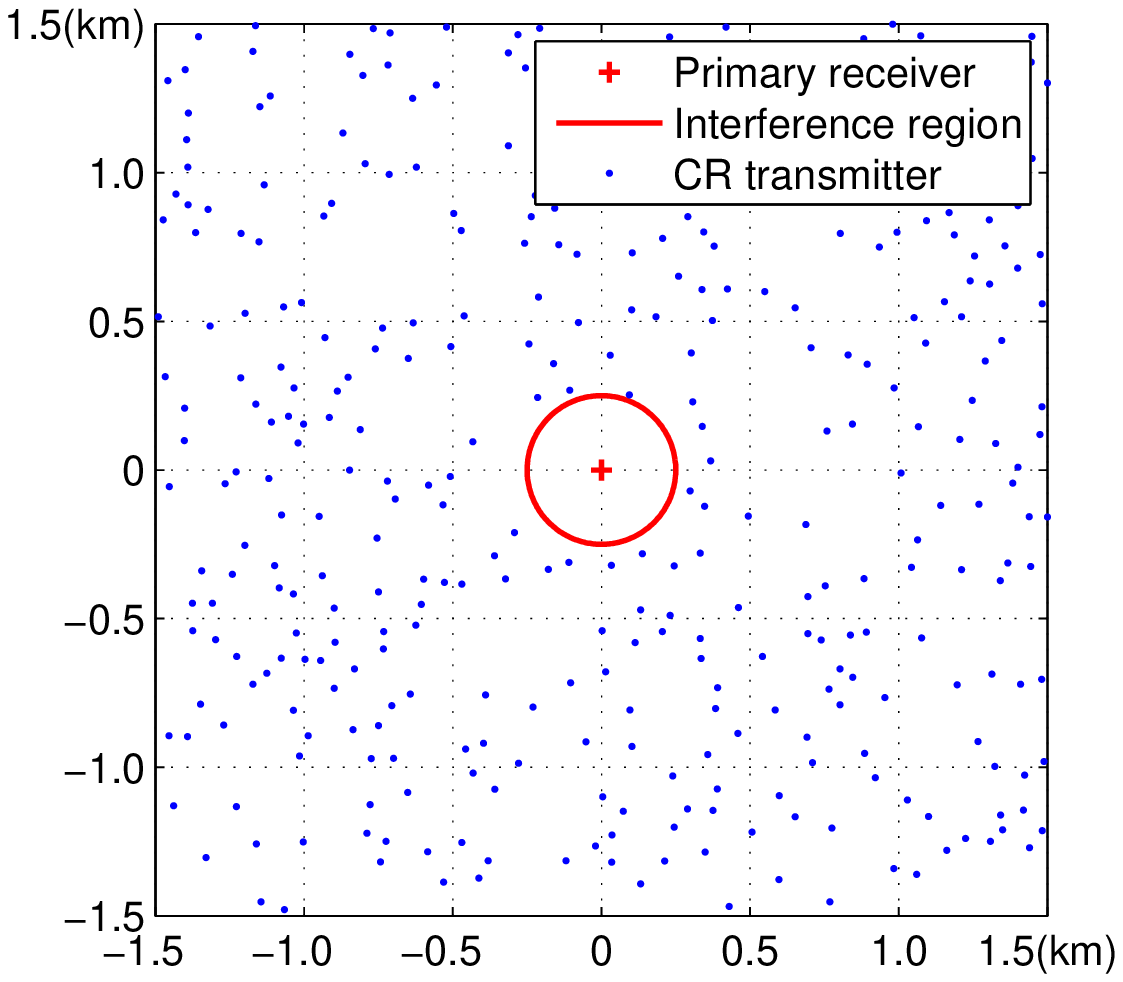} \\
\end{tabular}\\
\vspace*{-0.3cm}
\begin{center}
\small Fig.~1. System model for CR networks coexisting with a primary network ($R=250$ m).
\end{center}
\end{center}


\vspace*{1cm}
\begin{center}
\begin{tabular}{c}
\hskip-0.6cm\epsfxsize=10cm\epsffile{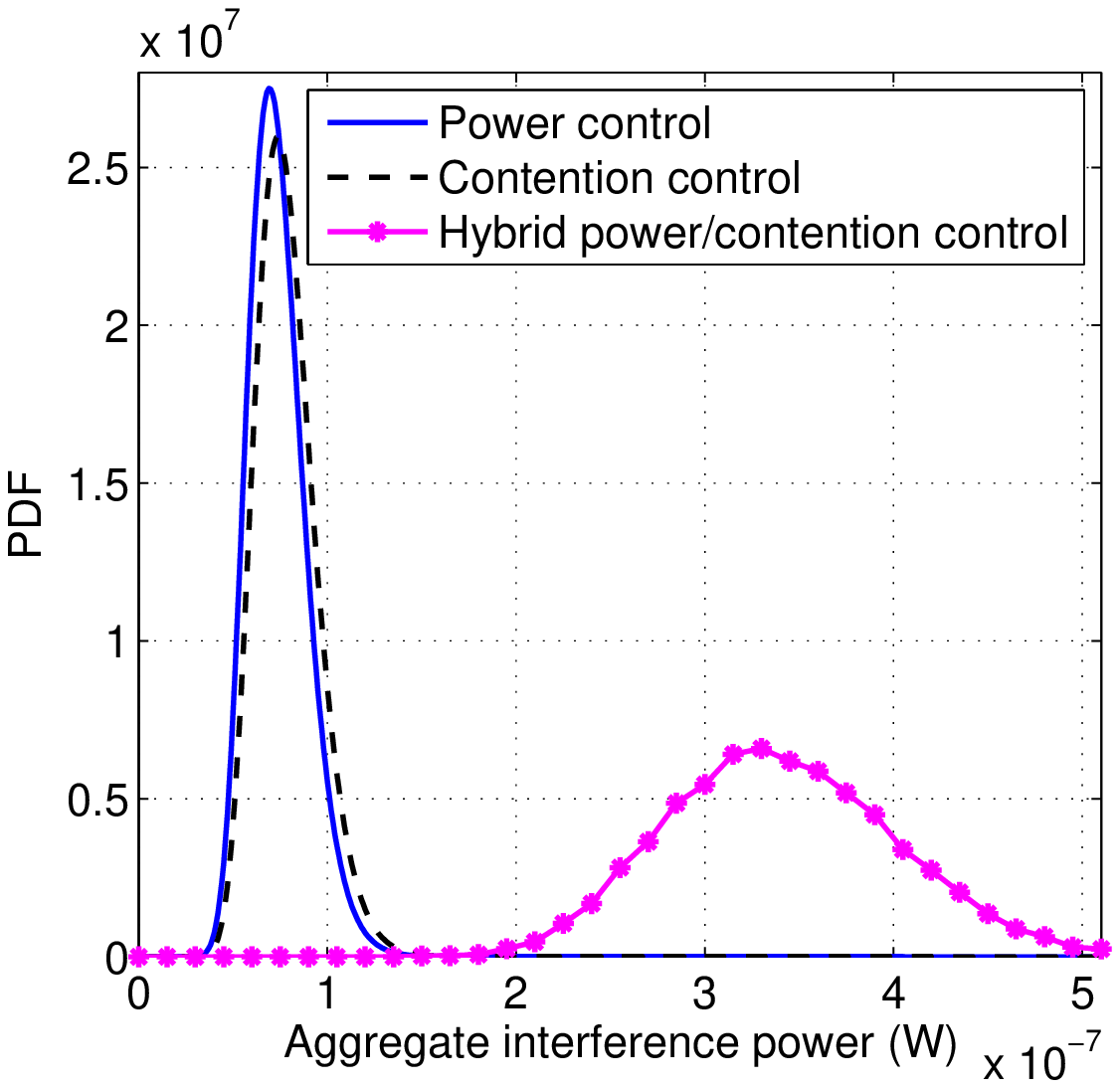} \\
\end{tabular}\\
\vspace*{-0.2cm}
\begin{center}
\small Fig.~2. Comparison of interference distributions for power, contention and hybrid power/contention control schemes ($R=$100 m, $\lambda=$3 user/$10^4$m$^2$,\ $\beta=$4, $r_\mathrm{pwc}=20$ m, $\alpha=4$, $P_\mathrm{max}=1$ W, $p=1$ W, $d_\mathrm{min}=20$ m and $r_\mathrm{hyb}=30~\rm{m}$).
\end{center}
\end{center}

\begin{center}
\begin{tabular}{c}
\hskip-0.6cm\epsfxsize=18cm\epsffile{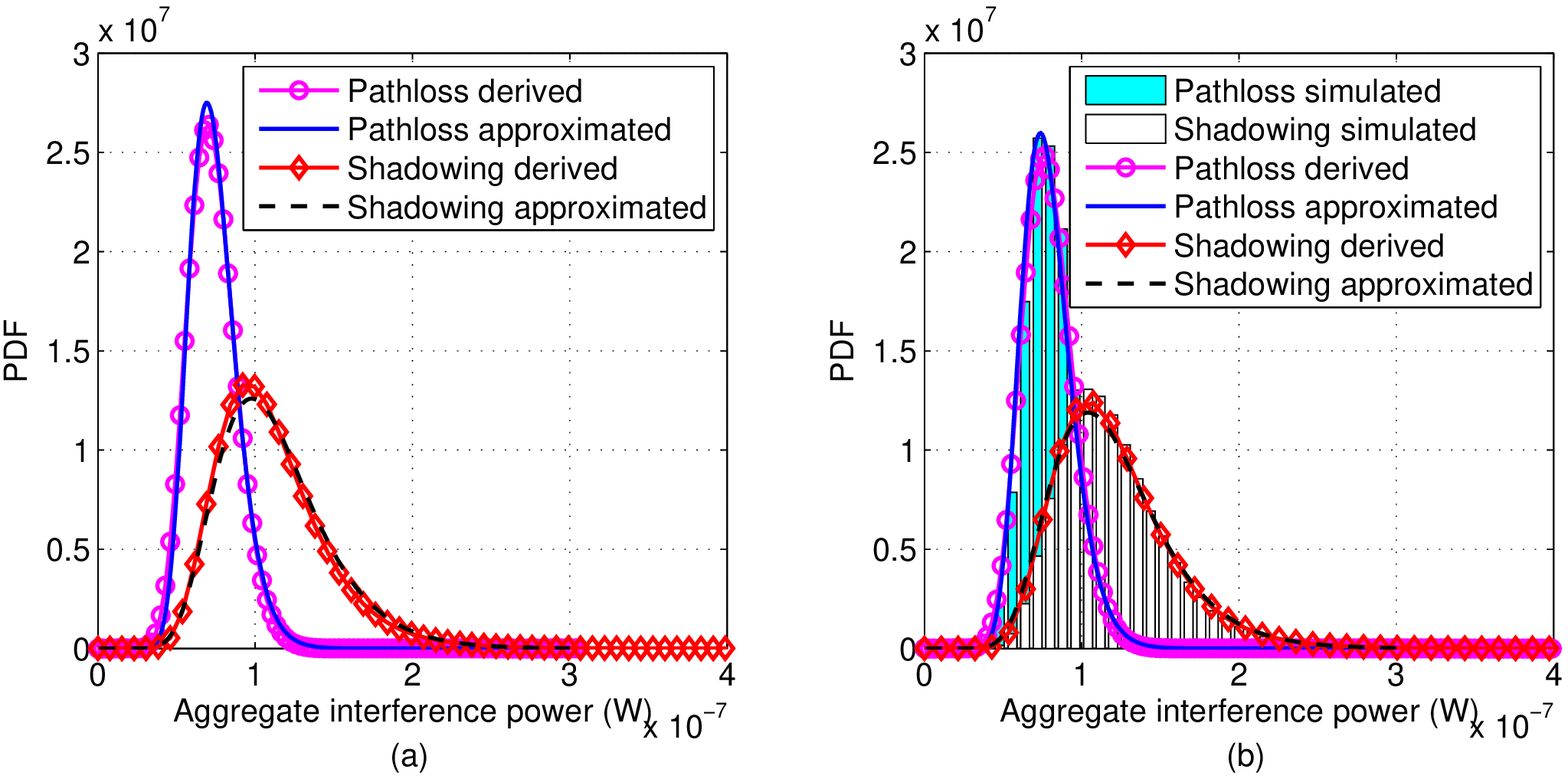} \\
\end{tabular}\\
\vspace*{-0.4cm}
\begin{center}
\small Fig.~3. Log-normal approximation for interference distribution under (a) power control ($R=$100 m, $\lambda=$3 user/$10^4$m$^2$,\ $\beta=$4, $r_\mathrm{pwc}=20$ m, $\alpha=4$, $P_\mathrm{max}=1$ W, $\mu=0$ and $\sigma=4$ dB) or (b) contention control ($R=$100 m, $\lambda=$3 user/$10^4$m$^2$,\ $\beta=$4, $d_\mathrm{min}=20$ m, $p=1$ W, $\mu=0$ and $\sigma=4$ dB).
\end{center}
\end{center}

\vspace*{1cm}
\begin{center}
\begin{tabular}{c}
\hskip-0.6cm\epsfxsize=9cm\epsffile{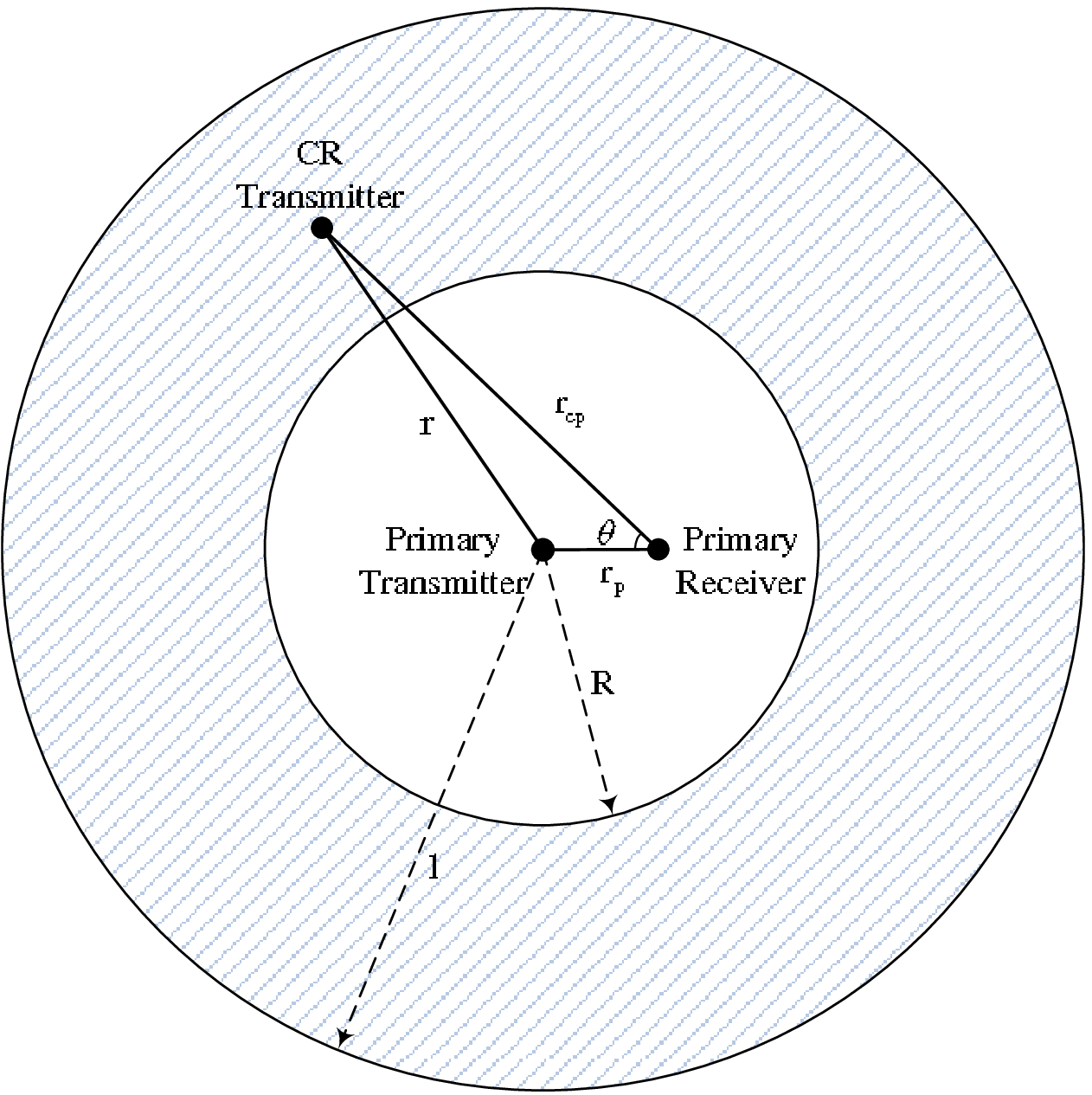} \\
\end{tabular}\\
\vspace{-0.3cm}
\begin{center}
\small Fig.~4. Imperfect knowledge of primary receiver location -  the primary receiver is hidden from all CR transmitters distributed in the shaded region.
\end{center}
\end{center}

\begin{center}
\begin{tabular}{c}
\hskip-0.6cm\epsfxsize=18cm\epsffile{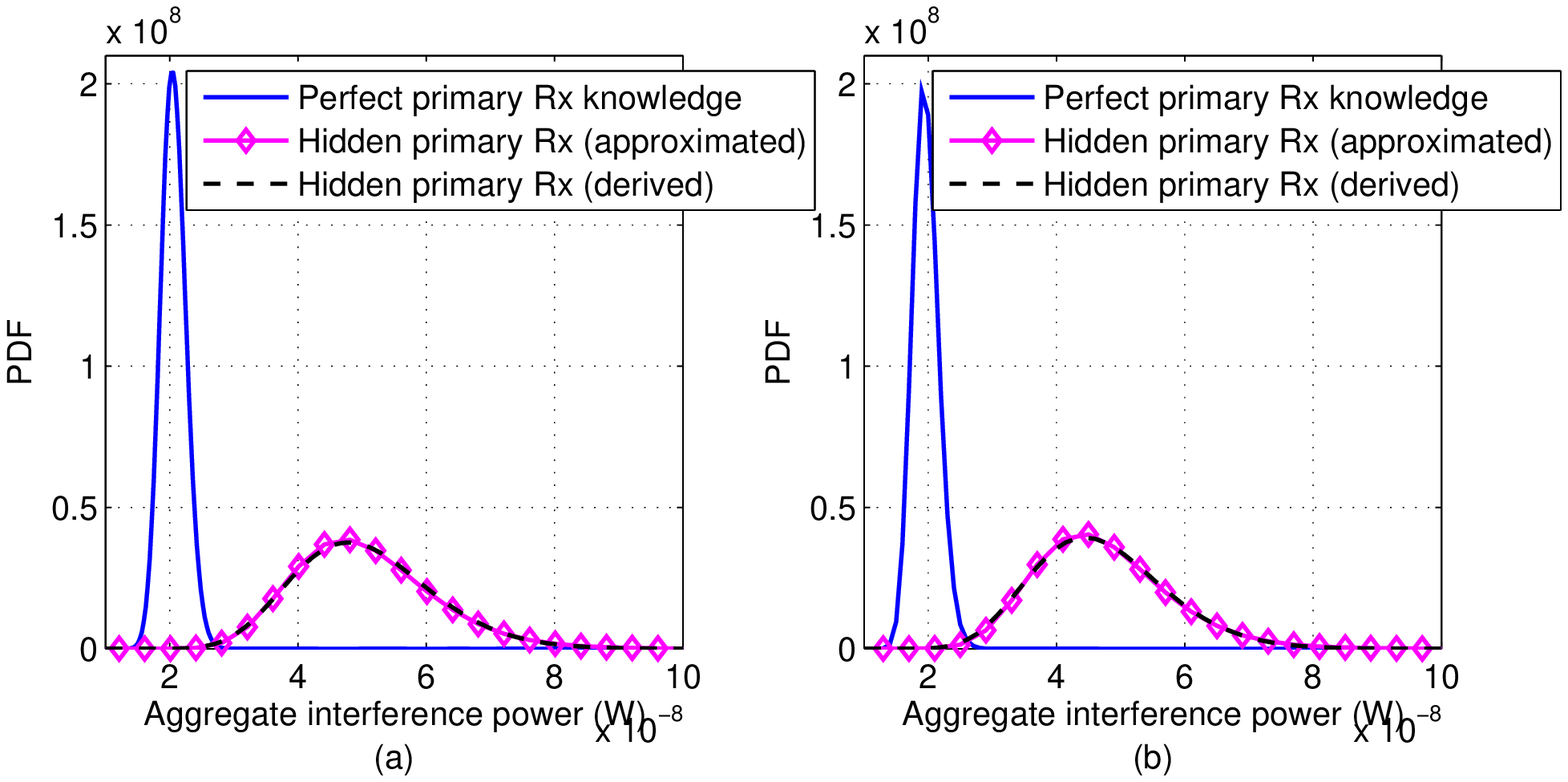} \\
\end{tabular}\\
\vspace{-0.4cm}
\begin{center}
\small Fig.~5. Log-normal approximation for interference distribution with a hidden primary receiver under (a)~power control ($R=$200 m, $\lambda=$3 user/$10^4$m$^2$,\ $\beta=$4, $r_\mathrm{pwc}=20$ m, $\alpha=4$, $P_\mathrm{max}=1$ W and $r_p=0.5R$) or (b)~contention control ($R=$200 m, $\lambda=$3 user/$10^4$m$^2$,\ $\beta=$4, $d_\mathrm{min}=20$ m, $p=1$ W and $r_p=0.5R$).
\end{center}
\end{center}

\vspace*{1cm}
\begin{center}
\begin{tabular}{c}
\hskip-0.6cm\epsfxsize=10cm\epsffile{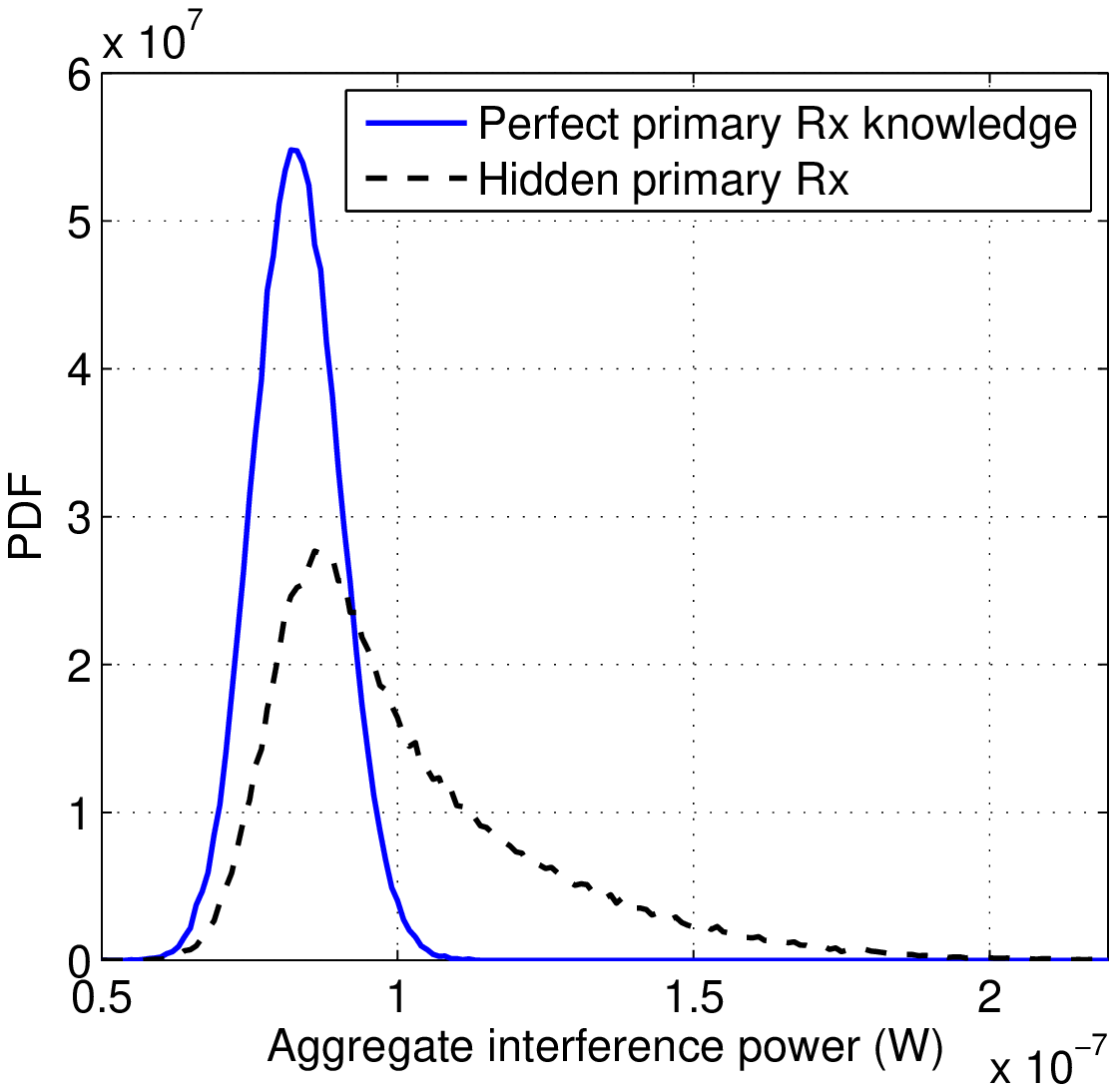} \\
\end{tabular}\\
\vspace{-0.3cm}
\begin{center}
\small Fig.~6. Impact of hidden primary receiver on interference distribution for CR networks under hybrid power/contention control scheme ($R=$200~m, $\lambda=$3 user/$10^4$m$^2$,\ $\beta=$4, $\alpha=4$, $d_\mathrm{min}=20$, $p=1$ W, $r_p=0.5R$ and $r_\mathrm{hyb}=30~\rm{m}$).
\end{center}
\end{center}


\begin{center}
\begin{tabular}{c}
\hskip-0.6cm\epsfxsize=18cm\epsffile{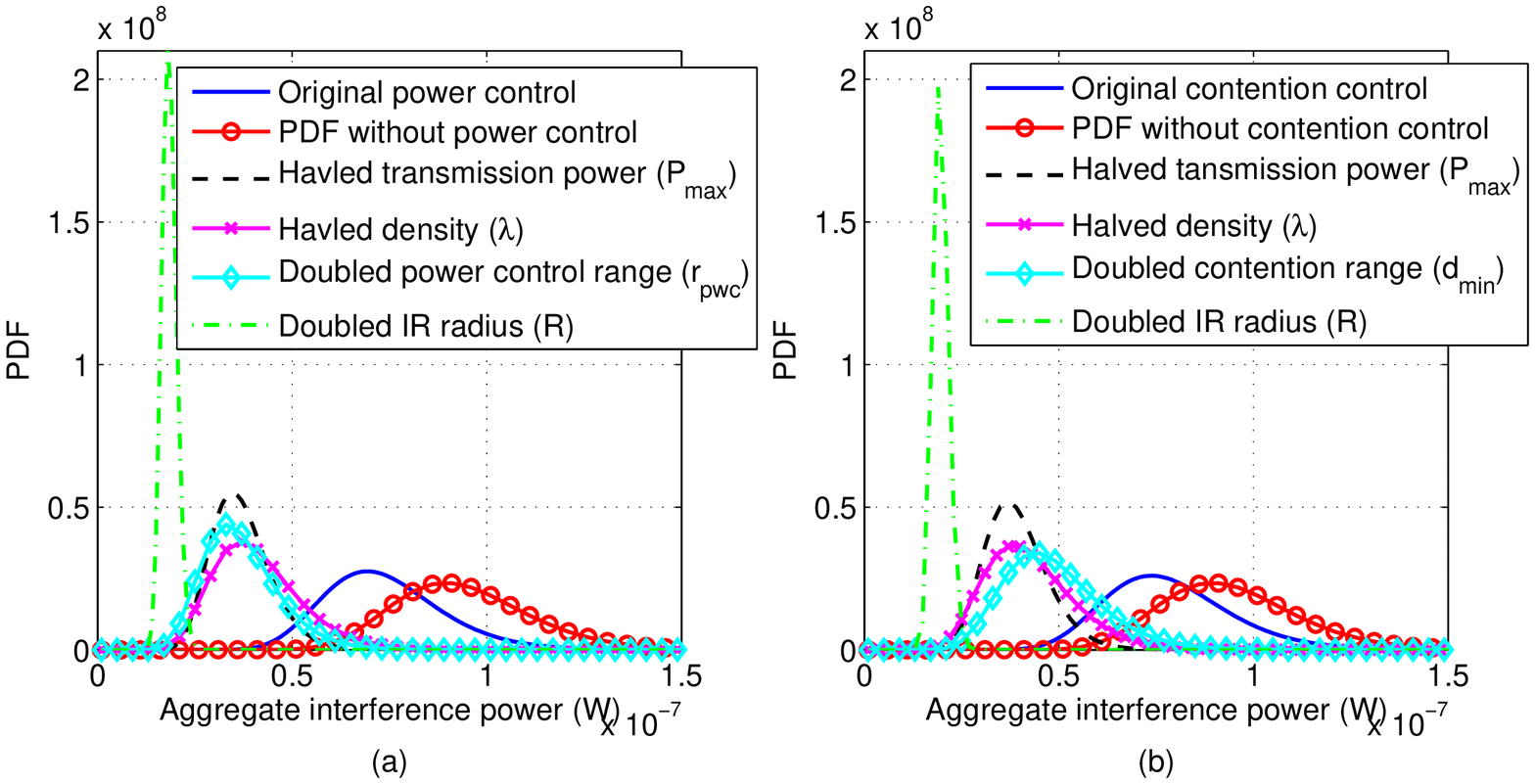} \\
\end{tabular}\\
\vspace{-0.4cm}
\begin{center}
\small Fig.~7. Impact of various CR deployment parameters on the aggregated interference for CR networks with (a)~power control ($R=$100 m, $\lambda=$3 user/$10^4$m$^2$,\ $\beta=$4, $r_\mathrm{pwc}=20$ m, $\alpha=4$ and $P_\mathrm{max}=1$ W) or (b)~contention control ($R=$100 m, $\lambda=$3 user/$10^4$m$^2$,\ $\beta=$4, $d_{\mathrm{min}}=$20 m, and $p=1$ W).
\end{center}
\end{center}

\vspace{1cm}
\begin{center}
\begin{tabular}{c}
\hskip-0.6cm\epsfxsize=18cm\epsffile{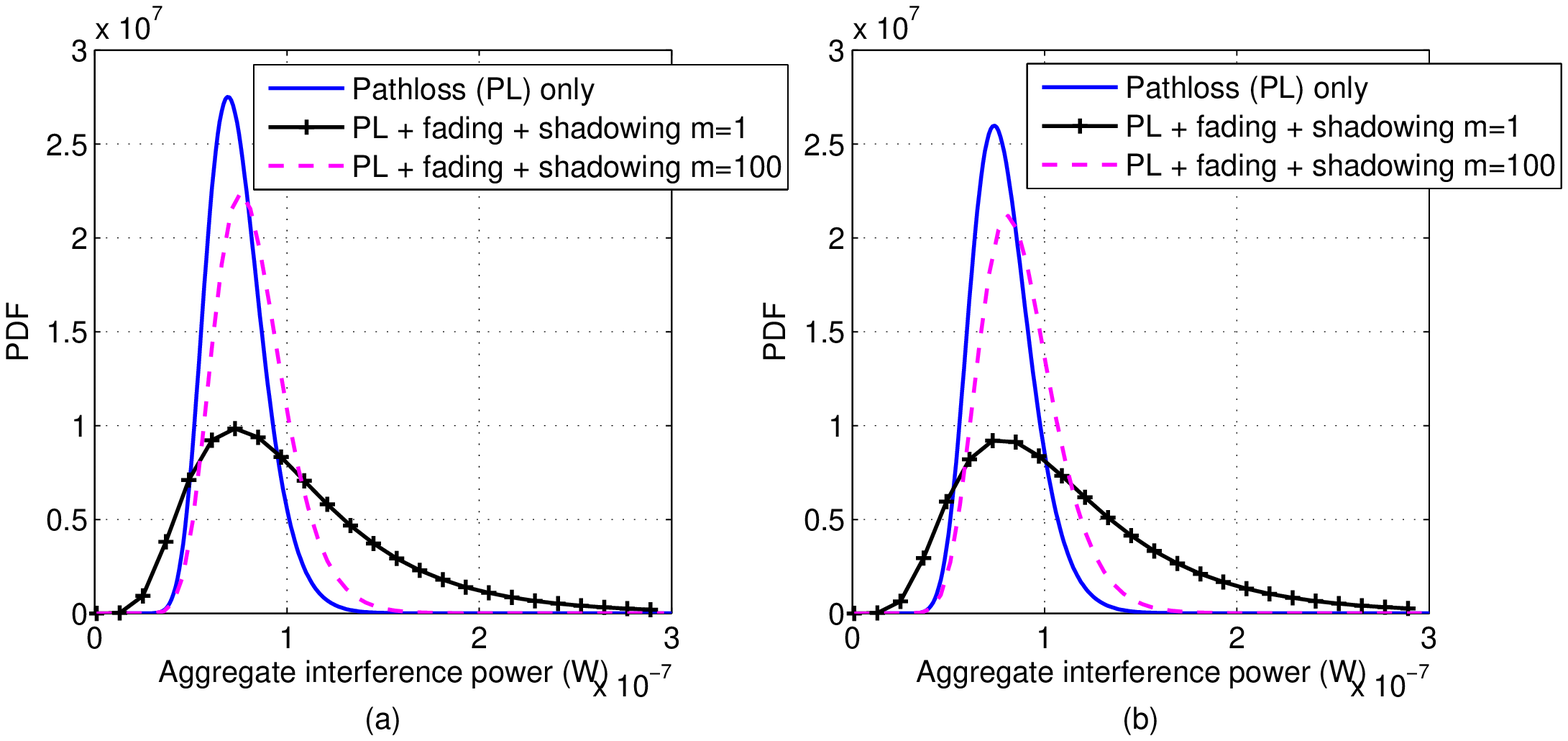} \\
\end{tabular}\\
\vspace{-0.4cm}
\begin{center}
\small Fig.~8. Impact of shadow fading on the aggregated interference for CR networks with (a)~power control ($R=$100 m, $\lambda=$3 user/$10^4$m$^2$,\ $\beta=$4, $r_\mathrm{pwc}=20$ m, $\alpha=4$ and $P_\mathrm{max}=1$ W) or (b)~contention control ($R=$100 m, $\lambda=$3 user/$10^4$m$^2$,\ $\beta=$4, $d_{\mathrm{min}}=$20 m and $p=1$ W).
\end{center}
\end{center}


\vspace{-0.1cm}
\begin{thebibliography}{1}

\bibitem{cr1}
S. Haykin, ``Cognitive radio: brain-empowered wireless communications,'' {\em IEEE J. Sel. Areas Commun.}, vol. 23, no. 2, pp.~201--220, Sept. 2005.

\bibitem{cr2}
I. F. Akyildiz, W. Y. Lee, M. C. Vuran, and S. Mohanty, ``NeXt generation/dynamic spectrum access/cognitive radio wireless networks: A survey,'' {\em Computer Networks}, vol. 50, no. 13, pp. 2127--2159, Sept. 2006.

\bibitem{cr3}
Q. Zhao and B. M. Sadler, ``A survey of dynamic spectrum access,'' {\em IEEE Signal Process. Mag.}, vol. 24, no. 3, 
pp. 79--89, May 2007.

\vspace{-0.05cm}
\bibitem{cr4}
C.-X. Wang, H.-H. Chen, X. Hong, and M. Guizani, ``Cognitive radio network management: tuning in to real-time conditions,'' {\em IEEE Veh. Technol. Mag.,} vol. 3, no. 1, pp. 28--35, Mar. 2008. 

\bibitem{cr5}
C.-X. Wang, X. Hong, H.-H. Chen, and J. S. Thompson, ``On capacity of cognitive radio networks with average interference power constraints,'' {\em IEEE Trans. Wireless Commun.,} vol. 8, no. 4, pp. 1620--1625, Apr. 2009. 

\bibitem{white}
X. Hong, C.-X. Wang, J. Thompson, and Y. Zhang, ``Demystifying white spaces,'' in {\em Proc. IEEE ICCCAS'08}, Xiamen, China, May 2008, pp. 350--354.

\bibitem{vt_magazine}
X. Hong, C.-X. Wang, H.-H. Chen, and Y. Zhang, ``Secondary spectrum access networks: recent development on the spatial models,'' {\em IEEE Veh. Technol. Mag.}, vol. 4, no. 2, pp. 36--43, June 2009.

\bibitem{cellular}
T. Kamakaris, D. Kivanc-Tureli, and U. Tureli, ``Interference model for cognitive coexistence in cellular systems,'' in {\em Proc. IEEE GLOBECOM'07}, Washington, DC, USA, Nov. 2007, pp. 4175--4179.

\bibitem{tv}
R. S. Dhillon and T. X. Brown, ``Models for analyzing cognitive radio interference to wireless microphones in TV bands,'' in {\em Proc. IEEE DySPAN'08}, Chicago, USA, Oct. 2008, pp. 1--10.

\bibitem{tv_gordon}
G. L. Stuber, S. Almalfouh, and D. Sale, ``Interference analysis of TV band whitespace,'' in {\em Proc. IEEE}, vol. 97, no. 4, pp. 741--754, Apr. 2009.

\bibitem{acc1}
N. Hoven and A. Sahai, ``Power scaling for cognitive radio,'' in {\em Proc. IEEE WNCMC'05}, Hawaii, USA, June 2005, pp.~250--255.

\vspace{-0.05cm}
\bibitem{acc2}
R. Menon, R. M. Buehrer and J. Reed, ``Outage probability based comparison of underlay and overlay spectrum sharing techniques,'' in {\em Proc. IEEE DySPAN'05}, Baltimore, USA, Nov. 2005, pp.~101--109.

\bibitem{acc3}
M. Timmers, S. Pollin, A. Dejonghe, A. Bahai, L. Van der Perre, and F. Catthoor, ``Accumulative interference modeling for cognitive radios with distributed channel access,'' in {\em Proc. IEEE CrownCom'08}, Singapore, May 2008.

\bibitem{xm}
X. Hong, C.-X. Wang, and J. S. Thompson, ``Interference modeling of cognitive radio networks,'' in {\em Proc. IEEE VTC'08-Spring}, Singapore, May 2008, pp.~1851--1855.

\bibitem{menon}
R. Menon, R. Buehrer, and J. Reed, ``On the impact of dynamic spectrum sharing techniques on legacy radio systems,'' {\em IEEE Trans. Wireless Commun.}, vol. 7, no. 11, pp. 4198--4207, Nov. 2008.

\bibitem{wcnc10}
Z. Chen, C.-X. Wang, X. Hong, J. Thompson, S. A. Vorobyov and X. Ge, ``Interference modeling for cognitive radio networks with power or contention control,'' in {\em Proc. IEEE WCNC 2010}, Sydney, Australia, Apr. 2010.

\bibitem{gordon}
G. L. Stuber, {\em Principles of Mobile Communication}, 2nd Edition, Boston: Kluwer Academic Publishers, 2001.

\bibitem{stochastic}
H. Q. Nguyen, F. Baccelli and D. Kofman, ``A stochastic geometry analysis of dense IEEE 802.11 networks,'' in {\em Proc. IEEE INFOCOM'07}, Anchorage, USA, May 2007, pp. 1199--1207.

\bibitem{pilot_sensing}
M. Ghosh, ``Text on FFT-based pilot sensing,'' IEEE 802.22, doc. no. 22-07-0298-01-0000, July 2007.

\bibitem{stoyan}
D. Stoyan, W. S. Kendall, and J. Mecke, {\em Stochastic Geometry and Its Applications}, Chichester: John Wiley \& Sons, 1986.

\bibitem{optimum90}
E. S. Sousa and J. A. Silvester, ``Optimum transmission range in a direct-sequence spread-spectrum multihop pack radio network,'' {\em IEEE J. Sel. Areas Commun.}, vol. 8, no. 5, pp. 762--771, June 1990.

\bibitem{poisson03}
X. Yang and A. P. Pertropulu, ``Co-channel interference modeling and analysis in a Poisson field of interferers in wireless communications,'' {\em IEEE Trans. Signal Process.}, vol. 51, no. 1, pp. 63--76, Jan. 2003.

\bibitem{poisson06}
P. C. Pinto and M. Z. Win, ``Communication in a poisson field of interferers,'' in {\em Proc. IEEE 40th Annual Conf. Inform. Sciences and Systems}, Princeton, USA, Mar. 2006, pp. 432--437.

\bibitem{paloheimo}
J. E. Paloheimo, ``On a theory of search,'' {\em Biometrika}, vol. 58, no. 1, pp. 61--75, Apr. 1971.

\bibitem{DStoyan}
D. Stoyan, ``On estimators of the nearest neighbour distance distribution function for stationary point processes,'' {\em Metrika}, vol. 64, no. 2, pp. 139-–150, Feb. 2006.

\bibitem{log-normal}
J. Salo, L. Vuokko, H. M. El-Sallabi, and P. Vainikainen, ``An additive model as a physical basis for shadow fading,'' {\em IEEE Trans. Veh. Technol.}, vol. 56, no. 1, pp.~13--26, Jan. 2007.

\bibitem{log-normal2}
M. Pratesi, F. Santucci, and F. Graziosi, ``Generalized moment matching for the linear combination of log-normal RVs: application\! to outage analysis in wireless systems,''\! {\em IEEE Trans. Wireless Commun.},\! vol. 5, no. 5, pp. 1122--1132, May 2006.

\bibitem{cumulant}
C. C. Chan and S. V. Hanly, ``Calculating the outage probability in a CDMA network with spatial poisson traffic,'' {\em IEEE Trans. Veh. Technol.}, vol. 50, no. 1, pp. 183--204, Jan. 2001.

\bibitem{logn_est}
R. Menon, R. M. Buehrer, and J. H. Reed, ``Impact of exclusion region and spreading in spectrum-sharing ad hoc networks,'' in {\em Proc. 1st Int. Workshop on Technology and Policy for Accessing Spectrum, TAPAS'06}, Aug. 2006.

\bibitem{logn_dis}
J. Aitchison and J. A. C. Brown, {\em The Lognormal Distribution}, Cambridge University Press, Cambridge UK, 1957. 


\bibitem{primary_detection}
B. Wild and K. Ramchandran, ``Detecting primary receivers for cognitive radio applications,'' in {\em Proc. IEEE DySPAN 2005}, Nov. 2005, pp. 124--130.

\end{thebibliography}
\end{document}